\documentclass[conference]{IEEEtran}

\usepackage{cite}
\usepackage{amsmath,amssymb,amsfonts}
\usepackage{amsthm}
\usepackage{csquotes}
\newtheorem{definition}{Definition}
\theoremstyle{definition}

\usepackage{algorithmic}
\usepackage{graphicx}
\usepackage{textcomp}
\usepackage{xcolor}
\usepackage{enumitem}
\usepackage{multirow}

\def\BibTeX{{\rm B\kern-.05em{\sc i\kern-.025em b}\kern-.08em
    T\kern-.1667em\lower.7ex\hbox{E}\kern-.125emX}}
    
\begin{document}

\title{A Layered Reference Model for Penetration Testing with Reinforcement Learning and Attack Graphs}

\author{\IEEEauthorblockN{Tyler Cody}
\IEEEauthorblockA{\textit{National Security Institute} \\
\textit{Virginia Tech}\\
Arlington, VA, USA \\
}
}

\maketitle



\begin{abstract}
This paper considers key challenges to using reinforcement learning (RL) with attack graphs to automate penetration testing in real-world applications from a systems perspective. RL approaches to automated penetration testing are actively being developed, but there is no consensus view on the representation of computer networks with which RL should be interacting. Moreover, there are significant open challenges to how those representations can be grounded to the real networks where RL solution methods are applied. This paper elaborates on representation and grounding using topic challenges of interacting with real networks in real-time, emulating realistic adversary behavior, and handling unstable, evolving networks. These challenges are both practical and mathematical, and they directly concern the reliability and dependability of penetration testing systems. This paper proposes a layered reference model to help organize related research and engineering efforts. The presented layered reference model contrasts traditional models of attack graph workflows because it is not scoped to a sequential, feed-forward generation and analysis process, but to broader aspects of lifecycle and continuous deployment. Researchers and practitioners can use the presented layered reference model as a first-principles outline to help orient the systems engineering of their penetration testing systems.
\end{abstract}


\section{Introduction}

\begin{figure}[t]
    \centering
    \includegraphics[width=0.45\textwidth]{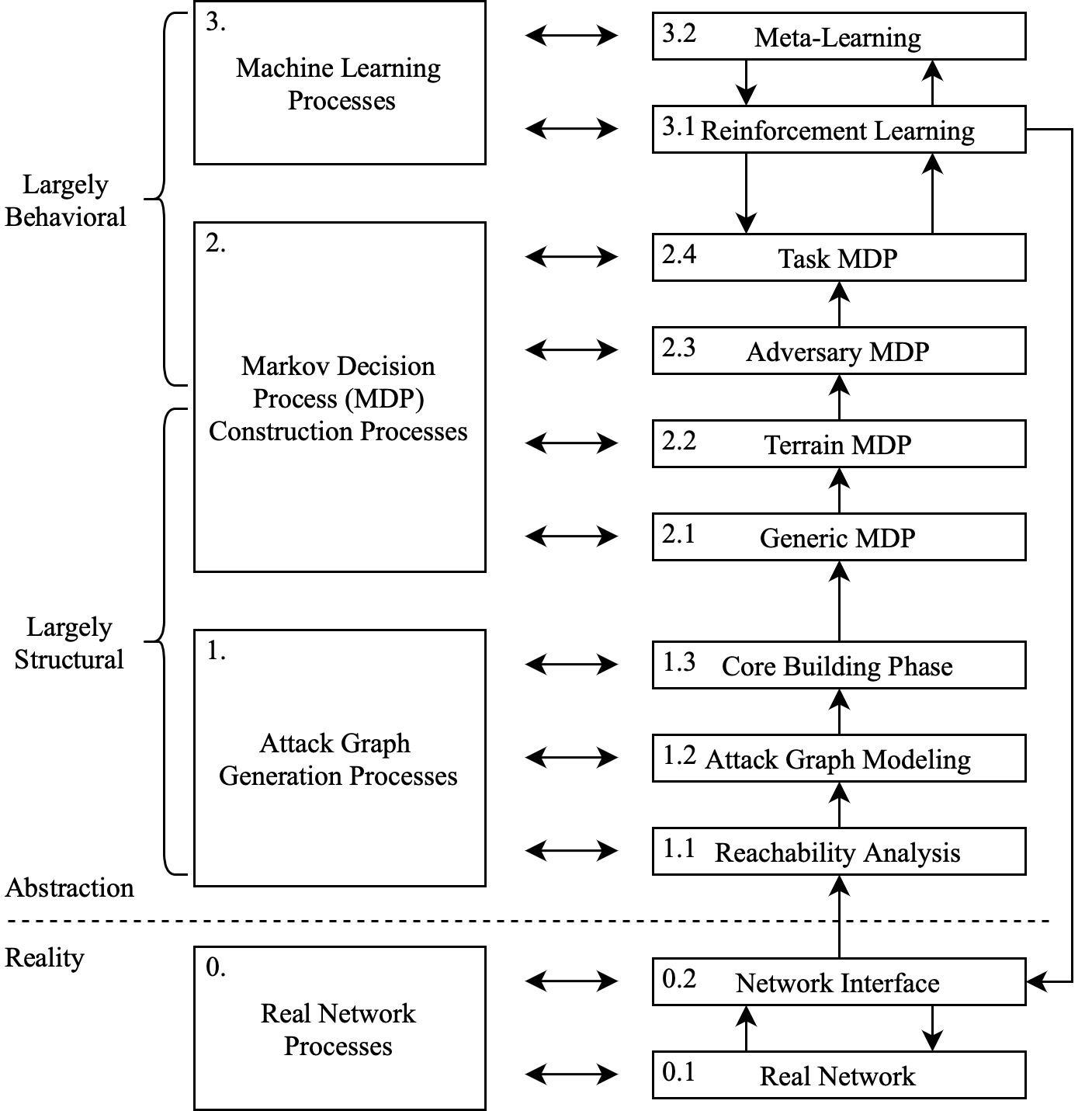}
    \caption{The presented layered reference model for RL with attack graphs (LRM-RAG). Structure and behavior refer to network (and path) structure and behavior.}
    \label{fig:mini-lrm}
\end{figure}

The challenge of real-world applications of artificial intelligence (AI) and machine learning (ML) is that they need representations of reality that are well-grounded to the system within which they operate. This is known as the \emph{grounding} problem \cite{harnad1990symbol, poggiolesi2021grounding}. The application of reinforcement learning (RL) to automate penetration testing is no different. Attack graphs are commonly used representations in the academic literature \cite{barik2016attack, kaynar2016taxonomy, zeng2019survey}. But how those attack graphs are actively grounded to real networks has been understudied in the literature from the perspective of RL. Leaving the grounding problem unaddressed puts the reliability of proposed RL-based cybersecurity solutions into serious question. This paper presents a layered reference model for how RL agents are grounded to real networks, shown in Fig. \ref{fig:mini-lrm}.

How exhaustive or succinct should attack graphs that RL agents interact with be? One concept of an attack graph is that it consists of all and only those paths from an initial state that reach terminal or success states, e.g., by violating a security or correctness property of a network model \cite{sheyner2002automated}. Exhaustive and succinct attack graphs like these, when used with RL, keep their exponential and polynomial growth rates (with respect to the number of hosts in a network) in order to turn the RL task into a needle-in-the-haystack problem of finding realistic attack paths or subgraphs within the graph. 

This bottlenecks the run-time of RL and directly violates a classic adage from learning theory, ``do not solve a specific problem by first solving a more general problem'' \cite{vapnik1999nature}. RL agents for penetration testing most often need to find a small set of realistic attack paths or subgraphs matching an attack template, playbook, adversary profile, etc.---to do so they do not need to find every possible attack path first. 

At the very least, formulating the RL agent's learning task as finding a subgraph within an exhaustive and succinct attack graph is an indirect formulation---the agent is more directly tasked with finding a or some realistic subgraphs within a network satisfying a goal (e.g., violating a security or correctness property of the tested network). The former assumes a full attack graph has already been generated and given to the agent, and, therefore, presupposes that generating such a graph is a necessary part of solution methods. The latter clearly does not.

It is not clear how the techniques developed and lessons learned from classical attack graph generation techniques like model checking \cite{jha2002two} and logic programming \cite{ou2006scalable} apply when RL agents, as opposed to human operators, will be interacting with the attack graphs. For example, should RL agents handle attack graph generation themselves? Or be confined to analysis? Or should they do both in an end-to-end manner? Answering these questions requires a broad view of the defining challenges of using RL with attack graphs. 

This paper provides a framing of these defining challenges in the form of a layered reference model termed LRM-RAG. The model is developed from a systems perspective rooted in the AI and ML character of using RL with attack graphs. 
LRM-RAG is defined and used to consider key challenges.
Challenges are organized into three questions:
\begin{itemize}
    \item How can RL interact with real networks in real-time?
    \item How can RL emulate realistic adversary behavior?
    \item How can RL deal with unstable/evolving networks?
\end{itemize}
These challenges are elaborated in Sections \ref{sec:generation}, \ref{sec:realism}, and \ref{sec:unstable}, respectively. When possible, the identified challenges are tied to ongoing research efforts in the RL with attack graphs literature.

\begin{figure}[t]
    \centering
    \includegraphics[width=0.45\textwidth]{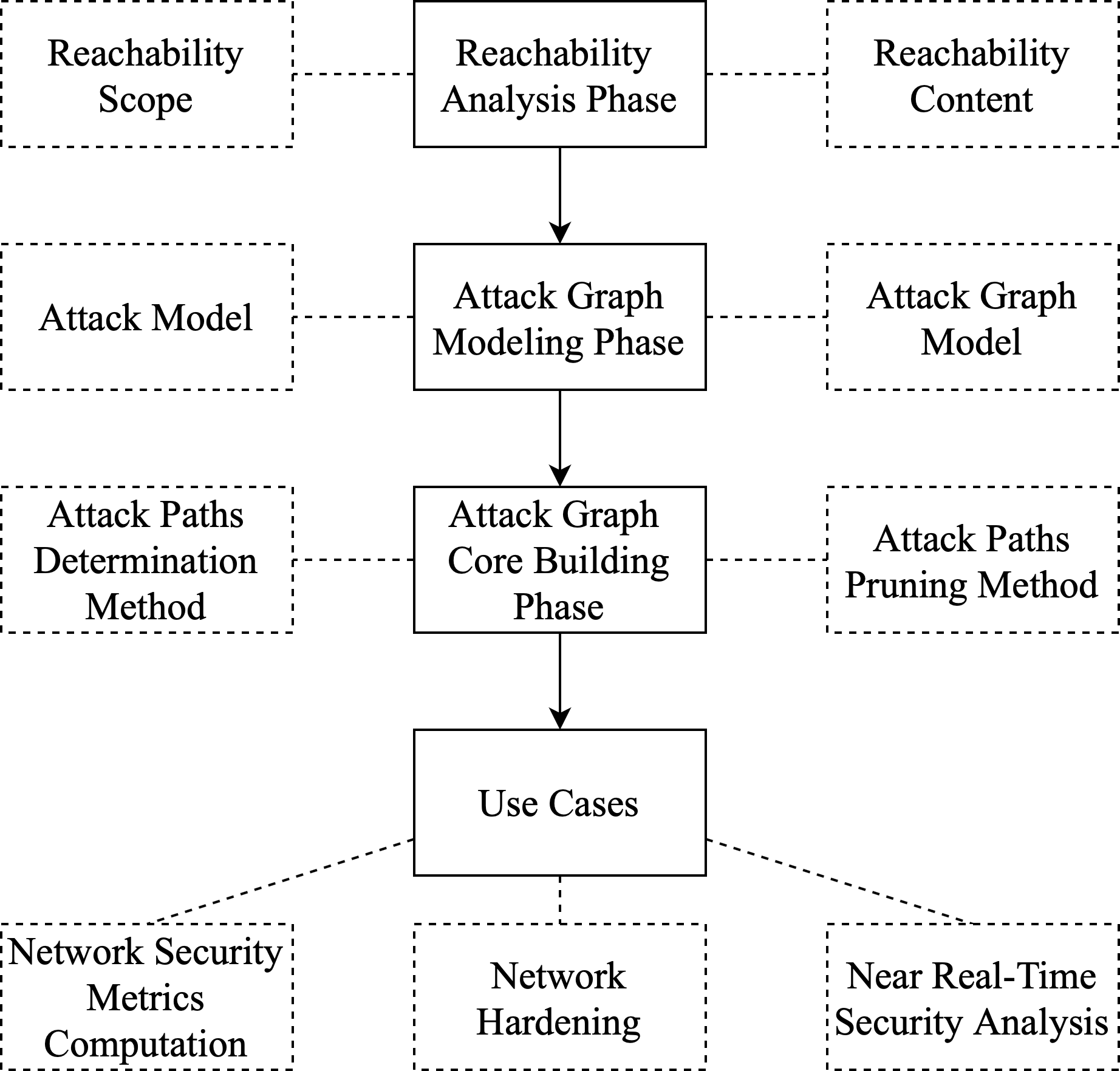}
    \caption{A taxonomy for attack graph generation and analysis proposed by Kaynar \cite{kaynar2016taxonomy}.}
    \label{fig:tax}
\end{figure}

Recent surveys of attack graph techniques provide a distillation of the major developments in attack graph generation \cite{barik2016attack}, a taxonomy for attack graph generation and analysis \cite{kaynar2016taxonomy}, and an overview of analysis methodologies \cite{zeng2019survey}. This paper is not a survey, however, it does offer an alternative, similarly broad perspective on attack graphs from the perspective of their use with RL. In this respect, this paper offers:
\begin{itemize}
    \item an extension of Kaynar's taxonomy for attack graph generation and analysis \cite{kaynar2016taxonomy}, shown in Figure \ref{fig:tax}, into Markov decision processes (MDPs) and RL, 
    \item an alternative, RL-oriented perspective to Barik et al.'s survey of attack graph generation methods \cite{barik2016attack},
    \item and a model for layering knowledge of terrain, adversaries, and tasks into generic MDPs over attack graphs.
\end{itemize}
In sum, this paper provides an accounting of the basic elements of using RL with attack graphs for practitioners.

The paper is structured as follows. First, related work is described in Section \ref{sec:related} and the RL for penetration testing literature is reviewed in Section \ref{sec:rl-pentesting}. Then, a layered reference model is defined that hierarchically (and visually) outlines penetration testing with RL and attack graphs in Section \ref{sec:layered}. Then, the key challenges are elaborated in Sections \ref{sec:generation}, \ref{sec:realism}, and \ref{sec:unstable}. The challenges all relate to questions of representation and grounding and they are organized into generation and actuation, realism, and unstable and evolving networks, respectively. The paper concludes in Section \ref{sec:conclusion} with a synopsis of the presented challenges and a discussion of the factors limiting researchers from addressing them.


\section{Related Work}
\label{sec:related}

\subsection{Attack Graph Generation}


Barik et al. \cite{barik2016attack} trace the development of attack graphs first from privilege graphs \cite{dacier1994privilege}, to state enumeration graphs \cite{phillips1998graph}, to scenario graphs \cite{sheyner2002automated}, then from there jointly to exploit dependency graphs \cite{ammann2002scalable, ammann2005host}, as used in topological vulnerability analysis (TVA), to logic programming graphs, as used in MulVal \cite{ou2006scalable}, and to multiple-prerequisite graphs, as used in NetSPA \cite{ingols2006practical}. Privilege graphs were a precursor to attack graphs that modeled users, their privileges, and their vulnerabilities \cite{dacier1994privilege}. Deemed too user-centric, state enumeration graphs which modeled hosts, their configurations, and their vulnerabilities directly were introduced \cite{phillips1998graph, sheyner2002automated}. But generating state enumeration graphs and the closely-related scenario graphs has an exponential growth rate with respect to the number of hosts in a network.

\begin{table*}[ht]
    \centering
    \caption{Attack Graph Generation Methods}
    \begin{tabular}{c|c|c|c}
        Method & Growth Rate in Hosts $n$ \cite{barik2016attack} & Are nodes hosts? & Is the graph acyclic? \\ \hline
        State-Enumeration~\cite{phillips1998graph} & $\mathcal{O}(2^n)$ & Yes & No \\
        Scenario~\cite{sheyner2002automated} & $\mathcal{O}(2^n)$ & Partially & Yes \\
        TVA~\cite{ammann2005host} & $\mathcal{O}(n^3)$ & No & Yes \\
        MulVal~\cite{ou2006scalable} & $\mathcal{O}(n^2)$ & No & Yes \\
        NetSPA~\cite{ingols2006practical} & $\mathcal{O}(n\log{}n)$ & Sometimes & No
    \end{tabular}
    \label{tab:generation-methods}
\end{table*}

By using the monotonicity assumption \cite{ammann2002scalable}, which specifies that privileges can only be gained, TVA approaches using exploit dependency graphs \cite{ammann2005host} and logic programming approaches like MulVal \cite{ou2006scalable} are able to reduce the exponential growth rate to polynomial. But to do so, both model the vulnerabilities or the vulnerabilities on hosts directly, as opposed to the hosts themselves, and therefore do not embed or represent the entire network model. NetSPA is able to further reduce the growth rate to be log-linear and maintains the concept of host at the node-level in attack graphs \cite{ingols2006practical}. But, NetSPA generates cyclical graphs, in contrast to the acyclical graphs generated by TVA, MulVal, and (exhaustive and succinct) model checking approaches to state enumeration graphs. Table \ref{tab:generation-methods} summarizes the basic character of these varied approaches to attack graph generation.

There are a number of works in the literature that seek to extend the methods in Table \ref{tab:generation-methods}. Topics include addressing completeness \cite{ning2004building}, complexity \cite{homer2008improving, ingols2009modeling}, and what-if analyses \cite{williams2008garnet, chu2010visualizing}. Current research into attack graph generation focuses on alert-driven methodology. This includes approaches that weigh completeness against alerts \cite{hu2020attack} and others that minimize the use of a priori expert knowledge \cite{nadeem2021alert}. Alert-driven attack graph generation can greatly reduce the scope of attack graph generation to relevant parts of networks, and is a promising area of research. Current efforts are tied more closely to attack graph visualization than to automated penetration testing with RL, however.

The process of generating and analyzing attack graphs is complex and merits its own taxonomy. Kaynar provides a general taxonomy for the basic workflow of using attack graphs \cite{kaynar2016taxonomy}, as shown in Figure \ref{fig:tax}. The basic workflow is that \emph{reachability analysis} provides a network model, then an \emph{attack graph model} must be chosen, e.g., from the aforementioned, and then the attack graph must be generated according to the chosen attack graph model and reachability analysis. This latter phase is termed the \emph{core building phase}. The final phase is to perform analysis of the attack graph, i.e., to apply it to a use case. Kaynar breaks these phases into a number of subtypes, but even Kaynar admits that the taxonomy is unable to partition the literature, i.e., generation and analysis methods belong to multiple subtypes, as attack graph generation and analysis is a complex and technically detailed process.


\subsection{Critiques of MDPs with Attack Graphs}

The use of ML in cyber applications has broad, general challenges \cite{fraley2017promise, apruzzese2018effectiveness, soni2019use}. The use of RL with attack graphs has drawn specific criticism. Shmaryahu et al. well elaborate the difficulties with assuming that MDPs modeled over attack graphs are fully observable, as well as with treating such difficulties with partially observable MDPs (POMDPs) \cite{shmaryahu2016constructing}. In short, they prefer attack trees to attack graphs. Formally speaking, it is unclear that taking the more specific tree structure instead of a graph structure makes a significant difference regarding observability if the full attack graph or full attack tree is not provided to the RL agent as a given. Notions of observability will be discussed herein.

Other criticisms concern realism and the use of vulnerability metrics like the Common Vulnerability Scoring System (CVSS) \cite{johnson2016can, munaiah2016vulnerability}. Specifically, the use of the CVSS and other vulnerability databases helps automate and scale MDP construction, but, if vulnerability information is used exclusively, it provides a representation of networks that is highly biased towards known vulnerabilities \cite{wu2012cyber}. Gangupantulu et al. propose balancing this bias by introducing concepts of cyber terrain \cite{raymond2014key, matania2020continuous} into MDPs, as will be discussed herein \cite{gangupantulu2021using}.

\subsection{Ontology Based Approaches}

One alternative to dealing with the complexities and difficulties of attack graphs is to take an entirely different orientation to applying RL to computer networks. Ontology-based approaches to penetration testing \cite{undercofer2003target, simmonds2004ontology, abdoli2010attacks, iannacone2015developing} can avoid the use of attack graphs by using model-based RL. Agent-based ontologies can use a formal or semi-formal domain specific language to interact directly with software tools that in turn interact with networks \cite{chu2018penetration, chu2020ontology, maeda2021automating}. In essence, this changes network structure from an external input to an internal representation the RL agent must learn. Hybrid approaches can provide ontology-based RL agents with network structure as an input, e.g., via an attack graph, but, naturally, this reintroduces the challenges described herein.

\section{RL for Penetration Testing}
\label{sec:rl-pentesting}

Recently, enabled by advances in deep learning, RL has seen broad application in cyber \cite{nguyen2019deep}, and a revival of interest from the penetration testing community in particular \cite{yousefi2018reinforcement, schwartz2019autonomous, ghanem2020reinforcement, chowdhary2020autonomous, hu2020automated, gangupantulu2021using, gangupantulu2021crown, nguyen2021proposal, zhou2021autonomous, zennaro2020modeling, tran2021deep, cody2022discovering}. Recently published literature on penetration testing with RL and attack graphs is summarized in Table \ref{tab:pentesting-methods}. 

\begin{table*}[t]
    \centering
    \caption{Penetration Testing Using RL with Attack Graphs}
    \begin{tabular}{l|l|l|l|l|l|l}
        \multicolumn{1}{c|}{Authors} & \multicolumn{1}{c|}{Graph Type} & \multicolumn{1}{c|}{Network Size} & \multicolumn{1}{c|}{Model} & \multicolumn{1}{c|}{Task} & \multicolumn{1}{c|}{Use CVSS?} & \multicolumn{1}{c}{Use terrain?} \\ \hline
        Yousefi et al. \cite{yousefi2018reinforcement} & MulVal & 44 vertices, 43 edges & MDP & Pathing & Yes & No \\
        Schwartz \& Kurniawati \cite{schwartz2019autonomous} & Network Model & 50 hosts & MDP & Pathing & No & No \\
        Ghanem \& Chen \cite{ghanem2020reinforcement} & Network Model & 100 hosts & POMDP & Unspecified & Yes & Yes \\
        Chowdary et al. \cite{chowdhary2020autonomous} & MulVal & 109 vertices, edges unknown & MDP & Pathing & Yes & No \\
        Hu et al. \cite{hu2020automated} & MulVal & 44 vertices, 52 edges & MDP & Pathing & Yes & No \\
        Gangupantulu et al. \cite{gangupantulu2021using} & MulVal & 955 vertices, 2350 edges & MDP & Pathing & Yes & Yes \\
        Gangupantulu et al. \cite{gangupantulu2021crown} & MulVal & 1617 vertices, 4331 edges & MDP & Crown Jewel Analysis & Yes & Yes \\
        Nguyen et al. \cite{nguyen2021proposal} & Network Model & 1024 hosts & MDP & Pathing & No & No \\
        Zhou et al. \cite{zhou2021autonomous} & Network Model & 17 hosts & MDP & Pathing & Yes & Yes \\
        Zennaro and Erodi \cite{zennaro2020modeling} & Network Model & Not reported & MDP & Capture the Flag & No & Yes \\
        Tran et al. \cite{tran2021deep} & Network Model & 900 hosts & MDP & Pathing & No & No \\
        Cody et al. \cite{cody2022discovering} & Network Model & 26 hosts & MDP & Exfiltration & Yes & Yes
    \end{tabular}
    \label{tab:pentesting-methods}
\end{table*}

Denis et al. define penetration testing as, ``a simulation of an attack to verify the security of a system or environment to be analyzed ... through physical means utilizing hardware, or through social engineering'' \cite{denis2016penetration}. If port scanning is looking through binoculars at a building to identify entry points, penetration testing is having someone actually break into the building. Penetration testing and static analysis combine to make up vulnerability detection and analysis \cite{bacudio2011overview}, \cite{shah2015overview}, \cite{chess2004static}.

Historically, penetration testing models have taken the form of the flaw hypothesis model \cite{pfleeger1989methodology, weissman1995penetration}, attack trees \cite{salter1998toward, schneier1999attack}, and attack graphs \cite{mcdermott2001attack}. Each approach has its own strengths. But attack graphs are particularly well-suited for RL. In fact, the earliest use of RL with attack graphs may be the use of the value iteration algorithm by Sheyner et al. to do probabilistic reliability analysis in their seminal work on scenario graphs \cite{sheyner2002automated}.

In RL, an agent is (typically) considered to interact with an environment $\mathcal{E}$ over a discrete number of time-steps by selecting an action $a_t$ at time-step $t$ from the set of actions $A$. In return, the environment $\mathcal{E}$ returns to the agent a new state $s_{t+1}$ and reward $r_{t+1}$. Thus, the interaction between the agent and environment $\mathcal{E}$ can be seen as a sequence $s_1, a_1, s_2, a_2, ..., a_{t-1}, s_t$. When the agent reaches a terminal state, the process stops. 

When $\mathcal{E}$ is a finite MDP, it is a tuple $\langle S, A, \Phi, P, R \rangle$, where $S$ is a set of states, $A$ is a set of actions, $\Phi \subset S \times A$ is the set of admissible state-action pairs, $P:\Phi \times S \to [0, 1]$ is the transition probability function, and $R: \Phi \to \mathbb{R}$ is the expected reward function where $\mathbb{R}$ is the set of real numbers. $P(s, a, s')$ denotes the transition probability from state $s$ to state $s'$ under action $a$, and $R(s, a)$ denotes the expected reward from taking action $a$ in state $s$. The goal of learning is to find a policy $\pi \in \Pi$, where $\Pi:S \to A$, that maximizes the sum of discounted future rewards.

As noted in Table \ref{tab:pentesting-methods}, other than Ghanem and Chen \cite{ghanem2020reinforcement}, the authors in the RL with attack graphs literature use fully observable MDPs to model networks. Many authors use the CVSS to furnish their MDPs. Yousefi et al. provide the earliest work doing so in deep RL for penetration testing \cite{yousefi2018reinforcement}. Hu et al. extend the use of the CVSS by proposing to use exploitability scores weight rewards \cite{hu2020automated}. Gangupantulu et al. \cite{gangupantulu2021using, gangupantulu2021crown} and Cody et al. \cite{cody2022discovering} explicitly extend the methods of Hu et al. with concepts of terrain. Gangupantulu et al. advocate defining models of terrain in terms of the rewards and transition probabilities of MDPs, first in the case of firewalls as obstacles \cite{gangupantulu2021using}, then in the case of lateral pivots nearby key terrain \cite{gangupantulu2021crown}. Cody et al. apply these concepts to exfiltration \cite{cody2022discovering}. Other authors either handcraft the MDP or do not remark on how its components are estimated.

Many authors apply generic deep Q learning (DQN) \cite{mnih2013playing, mnih2015human} to solve point-to-point network traversal tasks, termed `pathing' in Table \ref{tab:pentesting-methods} \cite{yousefi2018reinforcement, schwartz2019autonomous, chowdhary2020autonomous, hu2020automated, gangupantulu2021using}. Typically the terminal state is unknown and solutions take the form of individual paths. Others develop domain-specific modifications for deep RL including the double agent architecture \cite{nguyen2021proposal}, a hierarchical action decomposition approach \cite{tran2021deep}, and various improvements to DQN termed NDSPI-DQN \cite{zhou2021autonomous}. Another line of research focuses on developing more specific penetration testing tasks than `pathing'. A number of authors define more specific tasks by reward engineering and other modifications to the MDP including formulations of capture the flag \cite{zennaro2020modeling}, crown jewel analysis \cite{gangupantulu2021crown}, and discovering exfiltration paths \cite{cody2022discovering}. Existing literature will be discussed in the context of LRM-RAG in Section \ref{sec:challenges}. 

\section{Layered Reference Model}
\label{sec:layered}


LRM-RAG is constructed from general to specific, as shown when reading left-to-right in Figure \ref{fig:mini-lrm}. At the most general, a distinction is made between reality and the abstractions used to model it. Reality corresponds to the real network where penetration testing is applied. Attack graphs, MDPs, and RL are abstractions that model those networks, and they model structural and behavioral phenomena.

Structure is modeled by generating an attack graph and constructing an MDP. Next, behavior is modeled by enriching the transition probabilities and reward in the MDP and by training RL agents to interact with it. Then, informed by interacting with MDPs, RL agents can interact with real networks or inform interactions with real networks. RL agents observe outcomes of their actions in real networks or those they inform as filtered by the processes producing MDPs.

LRM-RAG gives a heterarchical view of automating penetration testing with RL and attack graphs. It contrasts the hierarchical (feed-forward) view offered by Kaynar \cite{kaynar2016taxonomy} depicted in Figure \ref{fig:tax}. The elements of LRM-RAG are classified in Table \ref{tab:table2}, and in the following, each element is defined explicitly.

\begin{table*}[htbp]
\centering
\caption{Classification of Elements in LRM-RAG}
\begin{tabular}{l|l|l|l}
\multicolumn{1}{c|}{Reality} & \multicolumn{3}{|c}{Abstraction}  \\ \hline
\hline
\multicolumn{1}{c|}{Layer 1} & \multicolumn{1}{|c|}{Layer 2} & \multicolumn{1}{|c|}{Layer 3} & \multicolumn{1}{|c}{Layer 4} \\
\multicolumn{1}{c|}{\emph{Real Network Processes}} & \multicolumn{1}{c|}{\emph{Attack Graph Generation Processes}} & \multicolumn{1}{|c|}{\emph{MDP Construction Processes}} & \multicolumn{1}{|c}{\emph{Machine Learning Processes}} \\
\hline
1.1 Real Network & 2.1 Reachability Analysis & 3.1 Generic MDP & 4.1 Reinforcement Learning \\
1.2 Network Interface & 2.2 Attack Graph Modeling & 3.2 Terrain MDP & 4.2 Meta-Learning \\
 & 2.3 Core Building Phase & 3.3 Adversary MDP & \\
 & & 3.4 Task MDP & \\
\end{tabular}
\label{tab:table2}
\end{table*}

\subsection{Real Network Processes}

The base layer in LRM-RAG is termed \emph{real network processes} and it consists of the real network and its interface.


\begin{definition} Real Network. \\ \emph{
The \emph{real network} is the (computer) network where penetration testing is applied.
}
\end{definition}

\noindent The real network may be uncertain, dynamic, and only partially observable. Defining the real network is more of a general scoping and place-setting activity than an exact specification of network properties. 



\begin{definition} Network Interface. \\ \emph{
The \emph{network interface} is the means for interaction with the real network.
}
\end{definition}

\noindent It has a two-way connection with the real network. It can be entirely automated or could involve cyber operators. 

\subsection{Attack Graph Generation Processes}

The next layer is termed \emph{attack graph generation processes}. It corresponds to Kaynar's taxonomy for generating attack graphs \cite{kaynar2016taxonomy}. Attack graphs are largely structural constructs that represent the topology of a network in consideration of vulnerabilities and notions of adversary success. 


\begin{definition} Reachability Analysis. \\ \emph{
\emph{Reachability analysis} discovers the basic topology of the network and produces a description of the connectivity between hosts and subnets.
}
\end{definition}




\begin{definition} Attack Graph Modeling. \\ \emph{
\emph{Attack graph modeling} selects an attack graph model, e.g., from Table \ref{tab:generation-methods}, that is appropriate for the use case. 
}
\end{definition}



\begin{definition} Core Building Phase. \\ \emph{
The \emph{core building phase} generates an attack graph corresponding to a selected model. 
}
\end{definition}
\noindent Kaynar can be consulted for more detail on attack graph generation \cite{kaynar2016taxonomy}.

\subsection{MDP Construction Processes}

The next layer is termed \emph{MDP construction processes}. Given a graphical model of a network, this layer adds additional structure (particularly in the form of terrain) and adds behavior (in the form of transition probabilities and rewards). Whereas the previous layer is meant to generate a generic state space for the real network, the MDP construction processes need to generate representations for the various penetration activities an RL agent is to perform. This means that there may be many specific MDPs associated with an attack graph. LRM-RAG treats the construction of these MDPs as a sequential process of layering detail on top of an attack graph. 


\begin{definition} Generic MDP. \\ \emph{
The \emph{generic MDP} is foundational for subsequent layers and its basic purpose is to represent the generic value (if any) of exploiting vulnerabilities and the likelihood of successfully exploiting vulnerabilities.
}
\end{definition}

\noindent Using the state space provided by attack graph generation, the generic MDP initializes rewards and transition probabilities for later layers to modify. A common approach is to use vulnerabilities databases, e.g., the CVSS, as a scalable approach to modeling behavior. 


\begin{definition} Terrain MDP. \\ \emph{
The \emph{terrain MDP} imbues the generic MDP with terrain, specifically, the kind of terrain uncovered through intelligence preparation of the battlefield.
}
\end{definition}

\noindent Terrain is general, meaning adversaries and tasks on a network share similar terrain. 


\begin{definition} Adversary MDP. \\ \emph{
The \emph{adversary MDP} scopes the RL agent's representation by tailoring the MDP to specific adversaries.
}
\end{definition}
\noindent Adversaries are modeled primarily by using attack templates, attack patterns, knowledge of red-team capabilities and infrastructure, etc. to define the set of actions an RL agent can take, but can also include varying transition probabilities with adversary skill or sophistication. 



\begin{definition} Task MDP. \\ \emph{
The \emph{task MDP} refines the adversary MDP to specify particular penetration testing activities.
}
\end{definition}

\noindent In general, the information specifying individual RL tasks is supplemental or otherwise not provided by the real network. A common effect of the task layer is to reallocate reward or to re-scope the state space. 


\subsection{Machine Learning Processes}

The final layer is termed \emph{machine learning processes}. At the highest level of abstraction, machine learning is used to approximate functions associated with penetration testing. RL is used to directly learn the MDP, while higher-level, meta-learning processes learn-to-learn. 


\begin{definition} Reinforcement Learning Agent \\ \emph{
The \emph{reinforcement learning agent} is an agent that maximizes the sum of discounted, expected future rewards in a MDP modeled over attack graphs.
}
\end{definition}

\noindent The RL agent interacts with the task MDP, and, the outcomes of that interaction are shared with the network interface. In effect, this creates two closed loops. The first, smaller loop is that between the RL agent and the task MDP. The second, much longer loop is that between the RL agent and real network. 
Transfer learning, multi-task learning, and other means of augmenting RL processes which are directly used to learn the MDP are at the same level of abstraction as the RL agent. Meta-learning occurs a level higher as its own learning process, and, naturally, can have it's own transfer learning, etc.

\begin{definition} Meta-Learning Agent. \\ \emph{
The \emph{meta-learning agent} is responsible for learning-to-learn new tasks in (new) networks.
}
\end{definition}

\noindent As tasks change, e.g., over the duration of an attack campaign, and as networks change, e.g., due to a dynamic user-base, new RL agents may need to be instantiated. Meta-learning takes a learning-to-learn approach to efficiently training new agents \cite{vanschoren2018meta}. 




\subsection{Summary}

\begin{figure*}[t]
    \centering
    \includegraphics[width=0.95\textwidth]{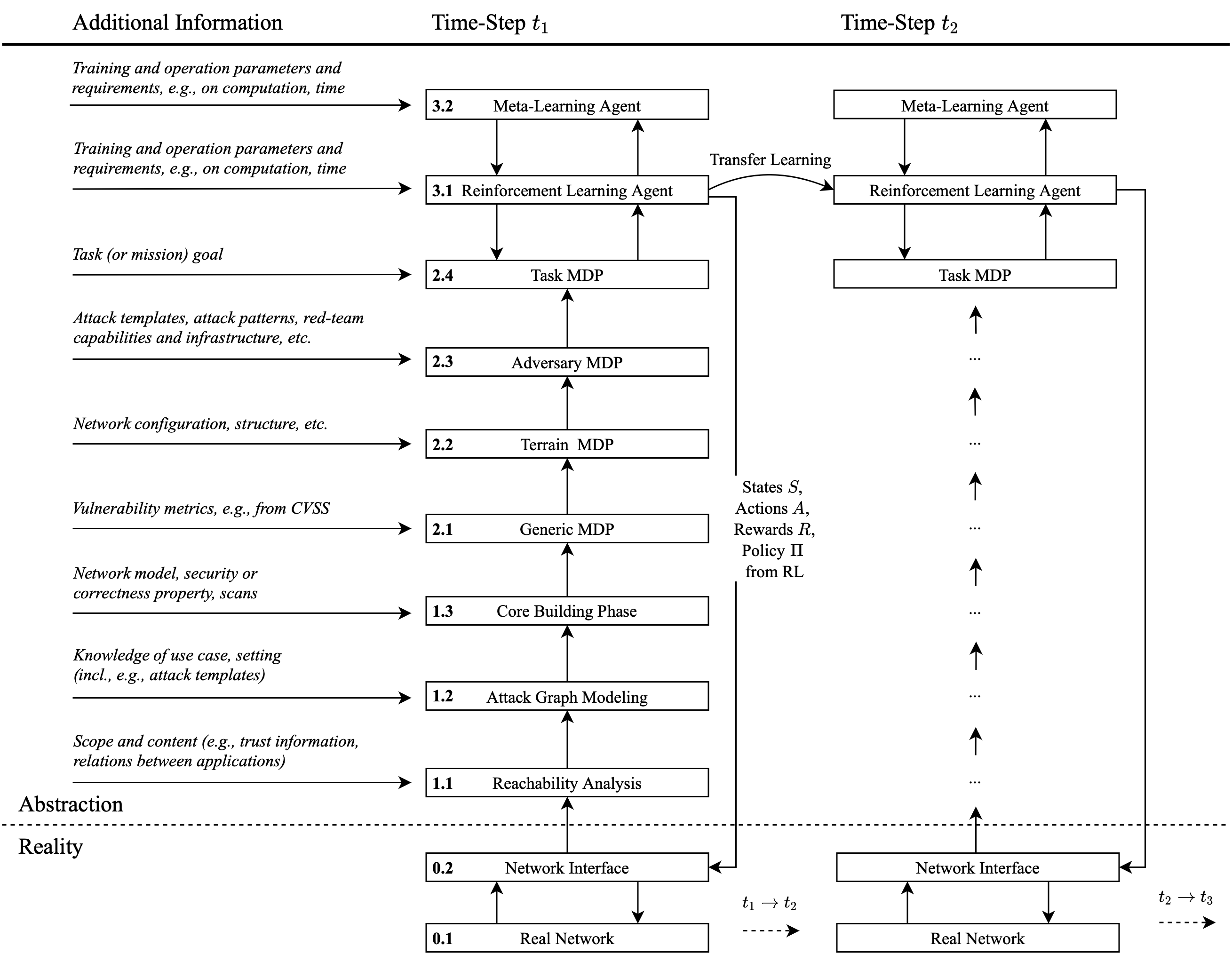}
    \caption{A layered reference model for automating penetration testing using RL and attack graphs with key additional information noted.}
    \label{fig:lrm}
\end{figure*}

LRM-RAG is depicted in Figure \ref{fig:lrm} and can be summarized as follows. Information from the real network is abstracted into an attack graph by the attack graph generation processes. This attack graph is a largely structural representation of vulnerabilities, their pre- and post-conditions, and (sometimes) hosts. To formulate an environment for the RL agent, a MDP is modeled over the generated attack graph using the MDP construction processes. The result of this is a layering of behavior and additional structure on top of the attack graph. The RL agent learns by interacting with the MDP, and as tasks, adversaries, and terrain change, the RL agent can transfer learn (i.e, share knowledge) or meta-learn (i.e., learn-to-learn) and thereby reapply lessons learned across penetration testing activities. The RL agent's interaction with the MDP is interpreted by the network interface to realize penetration testing and its outcomes in reality.

\subsection{Limitations}

As stated, the LRM-RAG is a reference model---not a survey. The view offered of RL and attack graphs by LRM-RAG does not model the specifics of penetration tasks. In the case that RL is treated as automating the general task of penetration testing, those tasks are essentially specifications on the MDP, i.e., they are at a lower-level of abstraction than LRM-RAG aims to model. Alternatively, in the case that RL is treated as automating specific penetration testing activities, it is not obvious what more can be generally said at a systems-level about specific tasks other than the basic layering of information involved in formulating them. Similarly, LRM-RAG does not survey or explicitly define state, action, and reward spaces. Formulations of MDPs for RL with attack graphs are a varied and evolving research topic. LRM-RAG identifies an emerging, general systems structure implicit to existing formulations, as evidenced by the connections drawn between existing literature and LRM-RAG in Section \ref{sec:challenges}.

\section{Key Challenges}
\label{sec:challenges}

The following subsections demonstrate the use of LRM-RAG for orienting engineers to key challenges to penetration testing with RL and attack graphs. First, generation and actuation is discussed. Then, realism is related to the MDP construction processes. Lastly, challenges in unstable and evolving networks are elaborated in terms of LRM-RAG.


\subsection{Generation and Actuation}
\label{sec:generation}

There are several challenges to generating attack graphs for RL agents and applying their policy or related analysis to real networks.
\begin{itemize}
    \item What is the role of supplemental information?
    \item How exhaustive and succinct an attack graph is needed?
    \item How much of the network model should be embedded into the attack graph?
\end{itemize}
\noindent LRM-RAG does not address these challenges, but offers a structured framework to support discussion.

The real network may be only partially observable, may be uncertain, and may be dynamic. The connection from the real network to the network interface is defining of what can be represented. However, LRM-RAG accounts for supplemental information being added to the RL agent's representation in subsequent steps, as depicted in Figure \ref{fig:lrm}. And so, some concern should be directed to the ratio of the information in the RL agent's representation directly observed through the network interface to the information supplemented in various subsequent layers.

There is not a universal value judgement that can be made as to whether this ratio should be higher or lower. It should be noted, though, that a low-latency or low-information connection from the real network to the network interface suggests an implicit requirement for a high-ratio of supplemental information. It can be noted, too, that a low-ratio of supplemental information suggests an implicit requirement for the attack graph generation, MDP construction, or RL agent to imbue the RL agent's representation with those concepts typically treated with supplemental information, e.g., terrain.

The character of supplemental information is coupled to the choice of attack graph model, and the choice of attack graph model is preceded by a choice of whether or not RL should be involved directly in the attack graph generation process. RL agents generating their own MDPs from reachability analysis may require a significant amount of representation learning to learn concepts like pre- and post-conditions, e.g., that seem essential to penetration testing. On the other hand, additional information needs to be layered into MDPs in the first place because of the narrow scope of highly optimized attack graph generation tools. These trade-offs are system, stakeholder, and use case specific.

The existing literature has taken two approaches to the generation challenge. As reported in Table \ref{tab:pentesting-methods}, Yousefi et al. \cite{yousefi2018reinforcement}, Chowdhary et al. \cite{chowdhary2020autonomous}, Hu et al. \cite{hu2020automated}, and Gangupantulu et al. \cite{gangupantulu2021using, gangupantulu2021crown} all use MulVal, while 
others use some form of network model---i.e., the network model is not processed into a graph of attack paths prior to interaction with RL. Instead, RL agents use network attack simulators or handcrafted network models. This divide seems to be tied to the role of supplemental information, as all authors using MulVal use the CVSS to provide vulnerability information.

MulVal, run to completion, will produce exhaustive and succinct attack graphs---attack graphs consisting of all and only the attack paths violating a given correctness or security property. The resulting environment $\mathcal{E}$ is therefore necessarily biased towards particular attacks and adversaries. Using network models as the environment $\mathcal{E}$ provides a kind-of generic template that can be refined using the MDP construction processes, and therefore without re-generating an (exhaustive and succinct) attack graph. While the scalability of retaining full network models, i.e., not processing them into proper attack graphs, in future networks is dubious, it is unclear if MulVal or other existing attack graph models are able to scale either.

The other connection is from the network interface to the real network. RL agents need a well-grounded means of observation as well as a well-grounded means of actuation. In LRM-RAG, the translation of the pattern of states, actions, and rewards generated by the RL agent into interactions with real networks is treated as a process internal to the network interface. This modeling decision---the decision to abstract away the interpretation of RL solutions to MDPs by path analysis programs or cyber operators---follows LRM-RAG's focus on RL. 

Significant post-processing of MDP solutions may mean greatly biasing information as it travels from the RL agent to the network. The literature underreports current practice for translating the interactions between attack graphs and RL into actions in the real network, but LRM-RAG suggests two loops grounding RL agents---a smaller loop between the agent and attack graph and a larger loop between the agent and the network interface. These two loops will be discussed in more detail in Section \ref{sec:unstable}.

\subsection{Realism}
\label{sec:realism}

Emulating realistic adversary behavior in networks is another key challenge to using RL with attack graphs. There are countless examples of RL agents solving MDPs by finding software bugs or modeling errors in their environment $\mathcal{E}$ which generate reward but do not lead the agent to completing its desired task. In RL with attack graphs, allotting reward for traversing a network switch undetected, e.g., could create positive feedback that incentivizes the agent to continuously loop through already exploited hosts to reuse the network switch. For this reason, reward engineering is a difficult and avoided task in RL.

LRM-RAG breaks MDP engineering into four parts. The generic MDP uses the CVSS to furnish incremental rewards for exploiting vulnerabilities and assigning transition probabilities \cite{yousefi2018reinforcement, chowdhary2020autonomous, hu2020automated, gangupantulu2021using}. Nearly half the authors in Table \ref{tab:pentesting-methods} implicitly use a generic MDP, however, even those that use it remark on its deficiencies. The terrain MDP addresses these deficiencies by introducing general concepts of terrain which apply to adversaries and tasks generally. Gangupantulu et al. \cite{gangupantulu2021using, gangupantulu2021crown} and Cody et al. \cite{cody2022discovering} propose several terrain concepts for RL with attack graphs including obstacles, key terrain, and cover and concealment.

Terrain may gives a realistic network behavior, perhaps, but terrain is not specific enough to give realistic adversary behavior. Adversaries have some preference over terrain, meaning different adversaries with different infrastructure, etc., will behave differently with the same terrain constraints. A nuanced challenge to modeling adversaries with MDPs is that campaigns are time-dependent. MDP engineering to represent an adversary's time-varying goals can treated using the task MDP. For example, the adversary MDP can set the action space based on an understanding of adversary capabilities, and then the task MDP can scope to a specific part of the network, modify terrain penalties, and set the terminal state \cite{gangupantulu2021crown, cody2022discovering}.

To an RL researcher, LRM-RAG appears to suggest extensive reward engineering. The convention in the RL literature is that reward engineering bottlenecks RL performance. However, there are a number of unavoidable steps to specifying penetration testing activities. These steps certainly lead to an MDP with engineered states, transitions, and rewards. And for rewards in particular, it is highly unlikely that the full scope of detail that a penetration testing system may be tasked with emulating can be summarized by allotting large, sparse terminal rewards to agents. Thus, LRM-RAG does not suggest reward engineering, but rather identifies a natural bias towards RL solutions that rely on extensive reward (and MDP) engineering. It is unclear that this natural bias can be removed without removing the meaning of penetration testing.

\subsection{Unstable and Evolving Networks}
\label{sec:unstable}

Recall the basic structure of LRM-RAG. Information from the real network is abstracted into an attack graph by the attack graph generation processes. The generated attack graph is a largely structural representation of vulnerabilities, their pre- and post-conditions, and (sometimes) hosts. To formulate an environment for the RL agent, a MDP is modeled over the generated attack graph using the MDP construction processes. The result of this is a layering of behavior and additional structure on top of the attack graph. The RL agent learns by interacting with the MDP, and as tasks, adversaries, and terrain change, the RL agent can transfer learn (i.e, share knowledge) or meta-learn (i.e., learn-to-learn) and thereby reapply lessons learned across penetration testing activities. The RL agent's interaction with the MDP is interpreted by the network interface to realize penetration testing and its outcomes in reality.

The long chain of functions between RL agents and real networks \emph{is} an RL agents grounding to reality. When information in an RL agent's representation is made available primarily by lower-level processes, the RL agent's representation is more directly grounded to the real network than when a majority of information comes from higher-level processes, e.g., from reward engineering tasks and terrain or from hand-engineering action sequences. But there is a dynamic aspect to the grounding problem as well. 

When the real network undergoes a change, that change is communicated to the machine learning processes by way of the network interface, attack graph generation processes, and MDP construction processes. Clearly, if the rate at which the real network changes is faster than the rate at which the RL agent's MDP can be created, then the RL agent will be grounded to an out-of-date representation of the real network while such a change occurs (and for a period thereafter). For largely static enterprise networks, or when analysis is offline, or when changes are small, this may not be a large concern. But in dynamic networks, a careful study should be made.

Recall the two loops identified in Section \ref{sec:generation}. In the smaller loop, the RL agent searches over attack graphs by way of interaction to find a particular subgraph or path. In the larger loop, this search process or discovered subgraph informs changes to the real network, e.g., it informs network hardening or resilience measures. In theory, these two loops could be similar length, if, e.g., the RL agent took actions in the attack graph at the same rate that the attack graph is refreshed. However, since once the task MDP is formed, its analysis is entirely computational, RL agent interaction with the MDP can likely, in general, be at least as fast. For example, the RL agent could interact every millisecond with an attack graph that is refreshed every 30 seconds.

Such a rate differential may make minute-to-minute host variations, e.g., due to virtual private network (VPN) use, a non-issue. But, in future networks, e.g., software-defined networks and Internet of Things networks \cite{bello2014intelligent, bekri2020softwarized}, the rate at which networks can evolve will seriously challenge automated penetration testing systems \cite{chen2018penetration}. Moreover, emulating phenomena like payload mutation and intelligent entry-point crawling may require learning under the assumption that networks are unstable and evolving, i.e., RL will be tasked with emulating adversary responses in changing networks.




\section{Conclusion}
\label{sec:conclusion}






The task of grounding RL agents to real networks is an open area of research. The choice to use attack graphs is not only a choice of (network) representation (for the RL agent), but also it is a specification on the grounding mechanism. Attack graphs are abstractions that must be generated, and that generation takes time. Consider, attack graphs (currently) grow at best log-linear and at worst exponentially with the number of hosts in the real network. Efforts to make attack graph generation computationally efficient highly bias representations, e.g., by not representing hosts, in effect, trading off the latency of the grounding mechanism for resolution in (network) representation. 

The information needed but left out by the generation process must be re-introduced. Customizable representations that can fit models of terrain or adversaries are highly desired in penetration testing activities, but relying on external information makes grounding tenuous. When a network destabilizes, changes, or otherwise evolves, detailed modeling becomes a bottleneck in re-establishing an RL agent's grounding. To that end, dynamic networks impose a preference for automated and empirical solution methods to constructing representations. In contrast, translating an RL agent's states and actions into interactions with real networks gives a nearly direct grounding. This asymmetry underpins LRM-RAG and is characteristic of its framing of continuous penetration testing with RL agents and attack graphs.

LRM-RAG provides a systems-level, extensible reference model for engineering the grounding of RL agents and their attack graphs to real networks. Domain experts can use LRM-RAG to give context to the individual engineering problems with which they are faced: attack graph generation, MDP formulation, RL algorithm design, etc. And, naturally, systems engineers can use LRM-RAG to relate those individual problems to each other.

\bibliographystyle{IEEEtran}
\bibliography{ref}

\begin{thebibliography}{10}
\providecommand{\url}[1]{#1}
\csname url@samestyle\endcsname
\providecommand{\newblock}{\relax}
\providecommand{\bibinfo}[2]{#2}
\providecommand{\BIBentrySTDinterwordspacing}{\spaceskip=0pt\relax}
\providecommand{\BIBentryALTinterwordstretchfactor}{4}
\providecommand{\BIBentryALTinterwordspacing}{\spaceskip=\fontdimen2\font plus
\BIBentryALTinterwordstretchfactor\fontdimen3\font minus
  \fontdimen4\font\relax}
\providecommand{\BIBforeignlanguage}[2]{{%
\expandafter\ifx\csname l@#1\endcsname\relax
\typeout{** WARNING: IEEEtran.bst: No hyphenation pattern has been}%
\typeout{** loaded for the language `#1'. Using the pattern for}%
\typeout{** the default language instead.}%
\else
\language=\csname l@#1\endcsname
\fi
#2}}
\providecommand{\BIBdecl}{\relax}
\BIBdecl

\bibitem{harnad1990symbol}
S.~Harnad, ``The symbol grounding problem,'' \emph{Physica D: Nonlinear
  Phenomena}, vol.~42, no. 1-3, pp. 335--346, 1990.

\bibitem{poggiolesi2021grounding}
F.~Poggiolesi, ``Grounding principles for (relevant) implication,''
  \emph{Synthese}, vol. 198, no.~8, pp. 7351--7376, 2021.

\bibitem{barik2016attack}
M.~S. Barik, A.~Sengupta, and C.~Mazumdar, ``Attack graph generation and
  analysis techniques.'' \emph{Defence Science Journal}, vol.~66, no.~6, 2016.

\bibitem{kaynar2016taxonomy}
K.~Kaynar, ``A taxonomy for attack graph generation and usage in network
  security,'' \emph{Journal of Information Security and Applications}, vol.~29,
  pp. 27--56, 2016.

\bibitem{zeng2019survey}
J.~Zeng, S.~Wu, Y.~Chen, R.~Zeng, and C.~Wu, ``Survey of attack graph analysis
  methods from the perspective of data and knowledge processing,''
  \emph{Security and Communication Networks}, vol. 2019, 2019.

\bibitem{sheyner2002automated}
O.~Sheyner, J.~Haines, S.~Jha, R.~Lippmann, and J.~M. Wing, ``Automated
  generation and analysis of attack graphs,'' in \emph{Proceedings 2002 IEEE
  Symposium on Security and Privacy}.\hskip 1em plus 0.5em minus 0.4em\relax
  IEEE, 2002, pp. 273--284.

\bibitem{vapnik1999nature}
V.~Vapnik, \emph{The nature of statistical learning theory}.\hskip 1em plus
  0.5em minus 0.4em\relax Springer Science \& Business Media, 1999.

\bibitem{jha2002two}
S.~Jha, O.~Sheyner, and J.~Wing, ``Two formal analyses of attack graphs,'' in
  \emph{Proceedings 15th IEEE Computer Security Foundations Workshop.
  CSFW-15}.\hskip 1em plus 0.5em minus 0.4em\relax IEEE, 2002, pp. 49--63.

\bibitem{ou2006scalable}
X.~Ou, W.~F. Boyer, and M.~A. McQueen, ``A scalable approach to attack graph
  generation,'' in \emph{Proceedings of the 13th ACM Conference on Computer and
  Communications Security}, 2006, pp. 336--345.

\bibitem{dacier1994privilege}
M.~Dacier and Y.~Deswarte, ``Privilege graph: an extension to the typed access
  matrix model,'' in \emph{European Symposium on Research in Computer
  Security}.\hskip 1em plus 0.5em minus 0.4em\relax Springer, 1994, pp.
  319--334.

\bibitem{phillips1998graph}
C.~Phillips and L.~P. Swiler, ``A graph-based system for network-vulnerability
  analysis,'' in \emph{Proceedings of the 1998 Workshop on New Security
  Paradigms}, 1998, pp. 71--79.

\bibitem{ammann2002scalable}
P.~Ammann, D.~Wijesekera, and S.~Kaushik, ``Scalable, graph-based network
  vulnerability analysis,'' in \emph{Proceedings of the 9th ACM Conference on
  Computer and Communications Security}, 2002, pp. 217--224.

\bibitem{ammann2005host}
P.~Ammann, J.~Pamula, R.~Ritchey, and J.~Street, ``A host-based approach to
  network attack chaining analysis,'' in \emph{21st Annual Computer Security
  Applications Conference (ACSAC'05)}.\hskip 1em plus 0.5em minus 0.4em\relax
  IEEE, 2005, pp. 10--pp.

\bibitem{ingols2006practical}
K.~Ingols, R.~Lippmann, and K.~Piwowarski, ``Practical attack graph generation
  for network defense,'' in \emph{2006 22nd Annual Computer Security
  Applications Conference (ACSAC'06)}.\hskip 1em plus 0.5em minus 0.4em\relax
  IEEE, 2006, pp. 121--130.

\bibitem{ning2004building}
P.~Ning, D.~Xu, C.~G. Healey, and R.~S. Amant, ``Building attack scenarios
  through integration of complementary alert correlation method.'' in
  \emph{NDSS}, vol.~4, 2004, pp. 97--111.

\bibitem{homer2008improving}
J.~Homer, A.~Varikuti, X.~Ou, and M.~A. McQueen, ``Improving attack graph
  visualization through data reduction and attack grouping,'' in
  \emph{International Workshop on Visualization for Computer Security}.\hskip
  1em plus 0.5em minus 0.4em\relax Springer, 2008, pp. 68--79.

\bibitem{ingols2009modeling}
K.~Ingols, M.~Chu, R.~Lippmann, S.~Webster, and S.~Boyer, ``Modeling modern
  network attacks and countermeasures using attack graphs,'' in \emph{2009
  Annual Computer Security Applications Conference}.\hskip 1em plus 0.5em minus
  0.4em\relax IEEE, 2009, pp. 117--126.

\bibitem{williams2008garnet}
L.~Williams, R.~Lippmann, and K.~Ingols, ``Garnet: A graphical attack graph and
  reachability network evaluation tool,'' in \emph{International Workshop on
  Visualization for Computer Security}.\hskip 1em plus 0.5em minus 0.4em\relax
  Springer, 2008, pp. 44--59.

\bibitem{chu2010visualizing}
M.~Chu, K.~Ingols, R.~Lippmann, S.~Webster, and S.~Boyer, ``Visualizing attack
  graphs, reachability, and trust relationships with navigator,'' in
  \emph{Proceedings of the Seventh International Symposium on Visualization for
  Cyber Security}, 2010, pp. 22--33.

\bibitem{hu2020attack}
H.~Hu, J.~Liu, Y.~Zhang, Y.~Liu, X.~Xu, and J.~Tan, ``Attack scenario
  reconstruction approach using attack graph and alert data mining,''
  \emph{Journal of Information Security and Applications}, vol.~54, p. 102522,
  2020.

\bibitem{nadeem2021alert}
A.~Nadeem, S.~Verwer, S.~Moskal, and S.~J. Yang, ``Alert-driven attack graph
  generation using s-pdfa,'' \emph{IEEE Transactions on Dependable and Secure
  Computing}, 2021.

\bibitem{fraley2017promise}
J.~B. Fraley and J.~Cannady, ``The promise of machine learning in
  cybersecurity,'' in \emph{SoutheastCon 2017}.\hskip 1em plus 0.5em minus
  0.4em\relax IEEE, 2017, pp. 1--6.

\bibitem{apruzzese2018effectiveness}
G.~Apruzzese, M.~Colajanni, L.~Ferretti, A.~Guido, and M.~Marchetti, ``On the
  effectiveness of machine and deep learning for cyber security,'' in
  \emph{2018 10th International Conference on Cyber Conflict (CyCon)}.\hskip
  1em plus 0.5em minus 0.4em\relax IEEE, 2018, pp. 371--390.

\bibitem{soni2019use}
S.~Soni and B.~Bhushan, ``Use of machine learning algorithms for designing
  efficient cyber security solutions,'' in \emph{2019 2nd International
  Conference on Intelligent Computing, Instrumentation and Control Technologies
  (ICICICT)}, vol.~1.\hskip 1em plus 0.5em minus 0.4em\relax IEEE, 2019, pp.
  1496--1501.

\bibitem{shmaryahu2016constructing}
D.~Shmaryahu, G.~Shani, J.~Hoffmann, and M.~Steinmetz, ``Constructing plan
  trees for simulated penetration testing,'' in \emph{The 26th International
  Conference on Automated Planning and Scheduling}, vol. 121, 2016.

\bibitem{johnson2016can}
P.~Johnson, R.~Lagerstr{\"o}m, M.~Ekstedt, and U.~Franke, ``Can the common
  vulnerability scoring system be trusted? a bayesian analysis,'' \emph{IEEE
  Transactions on Dependable and Secure Computing}, vol.~15, no.~6, pp.
  1002--1015, 2016.

\bibitem{munaiah2016vulnerability}
N.~Munaiah and A.~Meneely, ``Vulnerability severity scoring and bounties: Why
  the disconnect?'' in \emph{Proceedings of the 2nd International Workshop on
  Software Analytics}, 2016, pp. 8--14.

\bibitem{wu2012cyber}
J.~Wu, L.~Yin, and Y.~Guo, ``Cyber attacks prediction model based on bayesian
  network,'' in \emph{2012 IEEE 18th International Conference on Parallel and
  Distributed Systems}.\hskip 1em plus 0.5em minus 0.4em\relax IEEE, 2012, pp.
  730--731.

\bibitem{raymond2014key}
D.~Raymond, T.~Cross, G.~Conti, and M.~Nowatkowski, ``Key terrain in
  cyberspace: Seeking the high ground,'' in \emph{2014 6th International
  Conference On Cyber Conflict (CyCon 2014)}.\hskip 1em plus 0.5em minus
  0.4em\relax IEEE, 2014, pp. 287--300.

\bibitem{matania2020continuous}
E.~Matania and E.~Tal-Shir, ``Continuous terrain remodelling: gaining the upper
  hand in cyber defence,'' \emph{Journal of Cyber Policy}, vol.~5, no.~2, pp.
  285--301, 2020.

\bibitem{gangupantulu2021using}
R.~Gangupantulu, T.~Cody, P.~Park, A.~Rahman, L.~Eisenbeiser, D.~Radke, and
  R.~Clark, ``Using cyber terrain in reinforcement learning for penetration
  testing,'' \emph{arXiv preprint arXiv:2108.07124}, 2021.

\bibitem{undercofer2003target}
J.~Undercofer, A.~Joshi, T.~Finin, J.~Pinkston \emph{et~al.}, ``A
  target-centric ontology for intrusion detection,'' in \emph{Workshop on
  Ontologies in Distributed Systems, held at The 18th International Joint
  Conference on Artificial Intelligence}, 2003.

\bibitem{simmonds2004ontology}
A.~Simmonds, P.~Sandilands, and L.~Van~Ekert, ``An ontology for network
  security attacks,'' in \emph{Asian Applied Computing Conference}.\hskip 1em
  plus 0.5em minus 0.4em\relax Springer, 2004, pp. 317--323.

\bibitem{abdoli2010attacks}
F.~Abdoli, N.~Meibody, and R.~Bazoubandi, ``An attacks ontology for computer
  and networks attack,'' in \emph{Innovations and Advances in Computer Sciences
  and Engineering}.\hskip 1em plus 0.5em minus 0.4em\relax Springer, 2010, pp.
  473--476.

\bibitem{iannacone2015developing}
M.~Iannacone, S.~Bohn, G.~Nakamura, J.~Gerth, K.~Huffer, R.~Bridges,
  E.~Ferragut, and J.~Goodall, ``Developing an ontology for cyber security
  knowledge graphs,'' in \emph{Proceedings of the 10th Annual Cyber and
  Information Security Research Conference}, 2015, pp. 1--4.

\bibitem{chu2018penetration}
G.~Chu and A.~Lisitsa, ``Penetration testing for internet of things and its
  automation,'' in \emph{2018 IEEE 20th International Conference on High
  Performance Computing and Communications; IEEE 16th International Conference
  on Smart City; IEEE 4th International Conference on Data Science and Systems
  (HPCC/SmartCity/DSS)}.\hskip 1em plus 0.5em minus 0.4em\relax IEEE, 2018, pp.
  1479--1484.

\bibitem{chu2020ontology}
------, ``Ontology-based automation of penetration testing.'' in \emph{ICISSP},
  2020, pp. 713--720.

\bibitem{maeda2021automating}
R.~Maeda and M.~Mimura, ``Automating post-exploitation with deep reinforcement
  learning,'' \emph{Computers \& Security}, vol. 100, p. 102108, 2021.

\bibitem{nguyen2019deep}
T.~T. Nguyen and V.~J. Reddi, ``Deep reinforcement learning for cyber
  security,'' \emph{IEEE Transactions on Neural Networks and Learning Systems},
  2019.

\bibitem{yousefi2018reinforcement}
M.~Yousefi, N.~Mtetwa, Y.~Zhang, and H.~Tianfield, ``A reinforcement learning
  approach for attack graph analysis,'' in \emph{2018 17th IEEE International
  Conference On Trust, Security And Privacy In Computing And
  Communications/12th IEEE International Conference On Big Data Science And
  Engineering (TrustCom/BigDataSE)}.\hskip 1em plus 0.5em minus 0.4em\relax
  IEEE, 2018, pp. 212--217.

\bibitem{schwartz2019autonomous}
J.~Schwartz and H.~Kurniawati, ``Autonomous penetration testing using
  reinforcement learning,'' \emph{arXiv preprint arXiv:1905.05965}, 2019.

\bibitem{ghanem2020reinforcement}
M.~C. Ghanem and T.~M. Chen, ``Reinforcement learning for efficient network
  penetration testing,'' \emph{Information}, vol.~11, no.~1, p.~6, 2020.

\bibitem{chowdhary2020autonomous}
A.~Chowdhary, D.~Huang, J.~S. Mahendran, D.~Romo, Y.~Deng, and A.~Sabur,
  ``Autonomous security analysis and penetration testing,'' in \emph{2020 16th
  International Conference on Mobility, Sensing and Networking (MSN)}.\hskip
  1em plus 0.5em minus 0.4em\relax IEEE, 2020, pp. 508--515.

\bibitem{hu2020automated}
Z.~Hu, R.~Beuran, and Y.~Tan, ``Automated penetration testing using deep
  reinforcement learning,'' in \emph{2020 IEEE European Symposium on Security
  and Privacy Workshops (EuroS\&PW)}.\hskip 1em plus 0.5em minus 0.4em\relax
  IEEE, 2020, pp. 2--10.

\bibitem{gangupantulu2021crown}
R.~Gangupantulu, T.~Cody, A.~Rahma, C.~Redino, R.~Clark, and P.~Park, ``Crown
  jewels analysis using reinforcement learning with attack graphs,'' in
  \emph{2021 IEEE Symposium Series on Computational Intelligence (SSCI)}.\hskip
  1em plus 0.5em minus 0.4em\relax IEEE, 2021, pp. 1--6.

\bibitem{nguyen2021proposal}
H.~V. Nguyen, S.~Teerakanok, A.~Inomata, and T.~Uehara, ``The proposal of
  double agent architecture using actor-critic algorithm for penetration
  testing.'' in \emph{ICISSP}, 2021, pp. 440--449.

\bibitem{zhou2021autonomous}
S.~Zhou, J.~Liu, D.~Hou, X.~Zhong, and Y.~Zhang, ``Autonomous penetration
  testing based on improved deep q-network,'' \emph{Applied Sciences}, vol.~11,
  no.~19, p. 8823, 2021.

\bibitem{zennaro2020modeling}
F.~M. Zennaro and L.~Erdodi, ``Modeling penetration testing with reinforcement
  learning using capture-the-flag challenges: trade-offs between model-free
  learning and a priori knowledge,'' \emph{arXiv preprint arXiv:2005.12632},
  2020.

\bibitem{tran2021deep}
K.~Tran, A.~Akella, M.~Standen, J.~Kim, D.~Bowman, T.~Richer, and C.-T. Lin,
  ``Deep hierarchical reinforcement agents for automated penetration testing,''
  \emph{arXiv preprint arXiv:2109.06449}, 2021.

\bibitem{cody2022discovering}
T.~Cody, A.~Rahman, C.~Redino, L.~Huang, R.~Clark, A.~Kakkar, D.~Kushwaha,
  P.~Park, P.~Beling, and E.~Bowen, ``Discovering exfiltration paths using
  reinforcement learning with attack graphs,'' \emph{arXiv preprint
  arXiv:2201.12416}, 2022.

\bibitem{denis2016penetration}
M.~Denis, C.~Zena, and T.~Hayajneh, ``Penetration testing: Concepts, attack
  methods, and defense strategies,'' in \emph{2016 IEEE Long Island Systems,
  Applications and Technology Conference (LISAT)}.\hskip 1em plus 0.5em minus
  0.4em\relax IEEE, 2016, pp. 1--6.

\bibitem{bacudio2011overview}
A.~G. Bacudio, X.~Yuan, B.-T.~B. Chu, and M.~Jones, ``An overview of
  penetration testing,'' \emph{International Journal of Network Security \& Its
  Applications}, vol.~3, no.~6, p.~19, 2011.

\bibitem{shah2015overview}
S.~Shah and B.~M. Mehtre, ``An overview of vulnerability assessment and
  penetration testing techniques,'' \emph{Journal of Computer Virology and
  Hacking Techniques}, vol.~11, no.~1, pp. 27--49, 2015.

\bibitem{chess2004static}
B.~Chess and G.~McGraw, ``Static analysis for security,'' \emph{IEEE Security
  \& Privacy}, vol.~2, no.~6, pp. 76--79, 2004.

\bibitem{pfleeger1989methodology}
C.~P. Pfleeger, S.~L. Pfleeger, and M.~F. Theofanos, ``A methodology for
  penetration testing,'' \emph{Computers \& Security}, vol.~8, no.~7, pp.
  613--620, 1989.

\bibitem{weissman1995penetration}
C.~Weissman, ``Penetration testing,'' \emph{Information Security: An Integrated
  Collection of Essays}, vol.~6, pp. 269--296, 1995.

\bibitem{salter1998toward}
C.~Salter, O.~S. Saydjari, B.~Schneier, and J.~Wallner, ``Toward a secure
  system engineering methodolgy,'' in \emph{Proceedings of the 1998 Workshop on
  New Security Paradigms}, 1998, pp. 2--10.

\bibitem{schneier1999attack}
B.~Schneier, ``Attack trees,'' \emph{Dr. Dobb’s Journal}, vol.~24, no.~12,
  pp. 21--29, 1999.

\bibitem{mcdermott2001attack}
J.~P. McDermott, ``Attack net penetration testing,'' in \emph{Proceedings of
  the 2000 Workshop on New Security Paradigms}, 2001, pp. 15--21.

\bibitem{mnih2013playing}
V.~Mnih, K.~Kavukcuoglu, D.~Silver, A.~Graves, I.~Antonoglou, D.~Wierstra, and
  M.~Riedmiller, ``Playing atari with deep reinforcement learning,''
  \emph{arXiv preprint arXiv:1312.5602}, 2013.

\bibitem{mnih2015human}
V.~Mnih, K.~Kavukcuoglu, D.~Silver, A.~A. Rusu, J.~Veness, M.~G. Bellemare,
  A.~Graves, M.~Riedmiller, A.~K. Fidjeland, G.~Ostrovski \emph{et~al.},
  ``Human-level control through deep reinforcement learning,'' \emph{Nature},
  vol. 518, no. 7540, pp. 529--533, 2015.

\bibitem{vanschoren2018meta}
J.~Vanschoren, ``Meta-learning: A survey,'' \emph{arXiv preprint
  arXiv:1810.03548}, 2018.

\bibitem{bello2014intelligent}
O.~Bello and S.~Zeadally, ``Intelligent device-to-device communication in the
  internet of things,'' \emph{IEEE Systems Journal}, vol.~10, no.~3, pp.
  1172--1182, 2014.

\bibitem{bekri2020softwarized}
W.~Bekri, R.~Jmal, and L.~C. Fourati, ``Softwarized internet of things network
  monitoring,'' \emph{IEEE Systems Journal}, vol.~15, no.~1, pp. 826--834,
  2020.

\bibitem{chen2018penetration}
C.-K. Chen, Z.-K. Zhang, S.-H. Lee, and S.~Shieh, ``Penetration testing in the
  iot age,'' \emph{Computer}, vol.~51, no.~04, pp. 82--85, 2018.

\end{thebibliography}

\end{document}